\documentclass{article}
\usepackage{spconf,amsmath,amssymb,graphicx}
\usepackage{mathtools}
\usepackage{subfig}
\usepackage{siunitx}
\usepackage{bmpsize}
\usepackage[abs]{overpic}
\usepackage{adjustbox}
\usepackage{booktabs,array}
\usepackage{multirow}
\usepackage{balance}
\renewcommand{\thetable}{\Roman{table}}

\usepackage{accents}
\newlength{\dhatheight}
\newcommand{\doublehat}[1]{%
	\settoheight{\dhatheight}{\ensuremath{\hat{#1}}}%
	\addtolength{\dhatheight}{-0.35ex}%
	\hat{\vphantom{\rule{1pt}{\dhatheight}}%
		\smash{\hat{#1}}}}

\usepackage[usenames,dvipsnames]{color}
\usepackage[colorlinks=true,linkcolor=NavyBlue, citecolor=NavyBlue, urlcolor=NavyBlue]{hyperref}

\makeatletter
\renewcommand\footnotesize{%
	\@setfontsize\footnotesize\@ixpt{9.5}%
	\abovedisplayskip 8\p@ \@plus2\p@ \@minus4\p@
	\abovedisplayshortskip \z@ \@plus\p@
	\belowdisplayshortskip 4\p@ \@plus2\p@ \@minus2\p@
	\def\@listi{\leftmargin\leftmargini
		\topsep 4\p@ \@plus2\p@ \@minus2\p@
		\parsep 2\p@ \@plus\p@ \@minus\p@
		\itemsep \parsep}%
	\belowdisplayskip \abovedisplayskip
}
\makeatother

\title{Unsupervised adversarial correction of Rigid MR motion artifacts}
\name{Karim Armanious\textsuperscript{1,2}, Aastha Tanwar\textsuperscript{1}, Sherif Abdulatif\textsuperscript{ 1}, Thomas K\"ustner\textsuperscript{ 2,3}, Sergios Gatidis\textsuperscript{ 2}, Bin Yang\textsuperscript{1}}
\address{\textsuperscript{1}University~of~Stuttgart,~Institute~of~Signal~Processing~and~System~Theory,~Stuttgart,~Germany\\
	\textsuperscript{2}University~of~T\"ubingen,~Department~of~Radiology,~T\"ubingen,~Germany\\
	\textsuperscript{3}King's~College~London,~Biomedical~Engineering~Department,~London,~England
}
\begin{document}
%
\maketitle
\begin{abstract}

Motion is one of the main sources for artifacts in magnetic resonance (MR) images. It can have significant consequences on the diagnostic quality of the resultant scans. Previously, supervised adversarial approaches have been suggested for the correction of MR motion artifacts. However, these approaches suffer from the limitation of required paired co-registered datasets for training which are often hard or impossible to acquire. Building upon our previous work, we introduce a new adversarial framework with a new generator architecture and loss function for the unsupervised correction of severe rigid motion artifacts in the brain region. Quantitative and qualitative comparisons with other supervised and unsupervised translation approaches showcase the enhanced performance of the introduced framework. 
\end{abstract}
%
\section{Introduction}
\label{sec:intro}

Magnetic Resonance Imaging (MRI) is one of the most widely used medical imaging techniques. Its diagnostic capabilities have elevated it as a cornerstone of modern medical procedures. Nevertheless, due to long scanning times and the accompanying technical complexity, MR is susceptible to degradations in image quality due to motion artifacts. These degradations are dependant on the type and severity of the patient motion. Non-rigid artifacts are due to involuntary respiratory or cardiac motion which results in local deformations. Rigid artifacts are less subtle. They occur due to the bulk motion of body parts, e.g. the head or the legs, and prompt global deformations in the resultant scans.


There exists a large body of works which attempt to solve the problem of MR motion artifacts. Many of them focus on the prospective correction of the artifacts during the MR acquisition. For example, motion tracking devices such as cameras \cite{1}, respiratory belts \cite{2} and electrocardiogram \cite{3} have been used to trigger and guide the MR acquisition. This comes at the expense of the patients' comfort and longer scanning durations. Other prospective approaches employ motion-robust acquisition schemes \cite{4} and motion-resolved imaging \cite{5,6}. 

Also, retrospective correction of already acquired MR scans has been proposed as a solution for motion artifacts. For example, MR auto-focusing techniques have been utilized for the correction of complex motion artifacts \cite{7,8}. However, all of the above mentioned mechanisms rely on some a priori knowledge of the motion such as tracking signals or motion models

In recent years, deep learning has been utilized to support several medical tasks including MR motion correction. Convolutional neural networks (CNNs) were used to correct mild rigid motion artifacts \cite{9,10}. In our previous work, we proposed a generative adversarial network (GAN) framework \cite{11}, named MedGAN, which achieves state-of-the-art performance for the correction of severe motion artifacts in brain MRI \cite{12,13}. MedGAN was then extended for the correction of non-rigid motion artifacts in the abdomen and pelvis \cite{14}. However, this approach demands supervised training datasets which consist of paired motion-free and motion-corrupted MR scans. Such datasets require extensive planning and acquisition effort. Additionally, any misalignment in the training data has been reported to negatively affect the GAN output \cite{15}. To circumvent this drawback, we introduced in a recent work an unsupervised translation framework titled Cycle-MedGAN \cite{16}. This framework is trained using unpaired datasets without demanding any prior co-registration or alignment. However, Cycle-MedGAN was not capable yet of completely eliminating severe rigid motion artifacts. 

In this work, we extend our previous work on Cycle-MedGAN by incorporating several state-of-the-art advances from the deep learning field, such as self-attention \cite{17} and a new loss function \cite{18,19}. We apply the proposed framework on the task of correcting severe motion artifacts from brain MRI without utilizing any paired datasets during training while validating on a smaller subset of paired data. The performance of the proposed framework is compared against other unsupervised adversarial approaches qualitatively and quantitatively. Moreover, a comparison to a supervised translation approach is conducted to investigate the effects of training data misalignment on supervised GAN translation performance.

%
%
%

\begin{figure*}[!ht]
	\centering
	\includegraphics[width=0.87\textwidth]{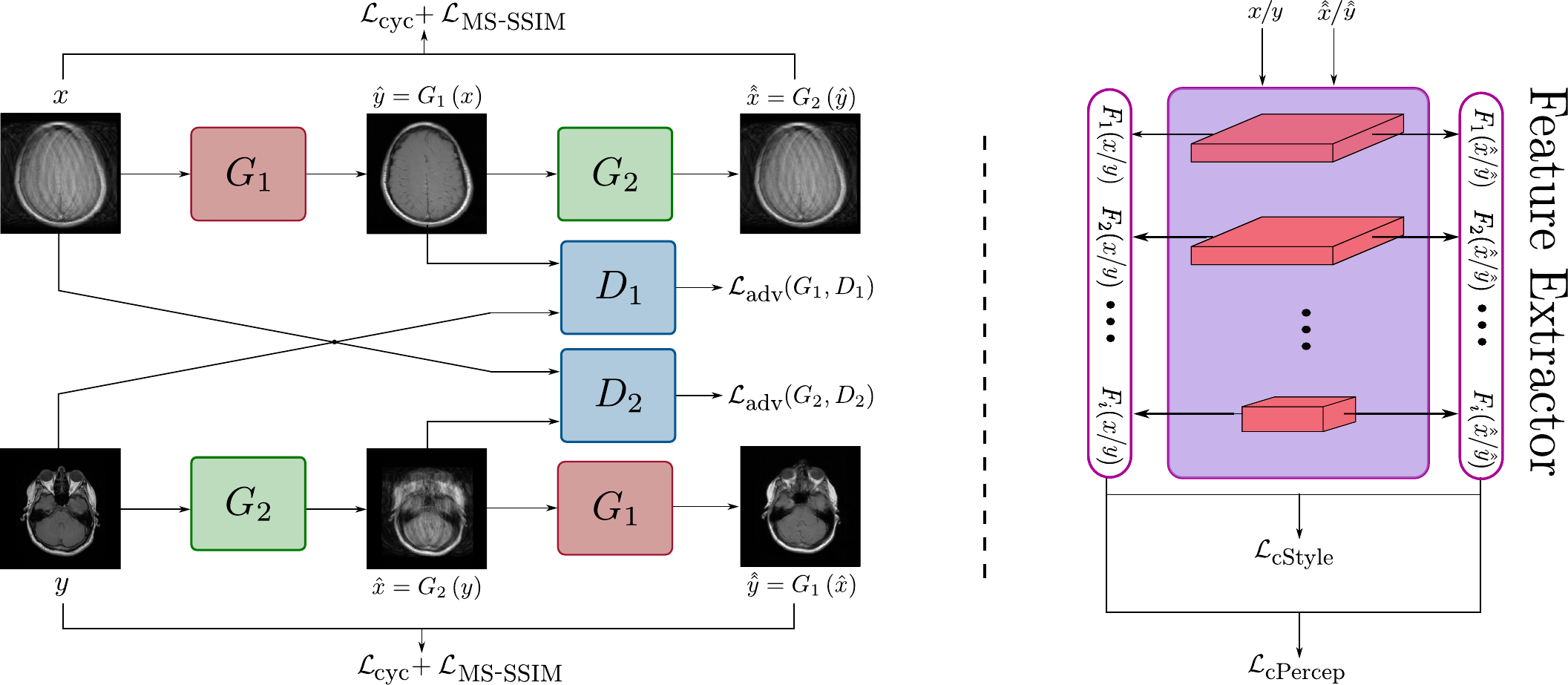}
	\caption{An overview of the improved Cycle-MedGAN framework with randomly sampled input images $x$ and $y$.}
	\label{1}
	\vspace{-4mm}
\end{figure*}

\section{Method}
\label{sec:format}
\vspace{-2mm}
In the following subsections, the proposed framework based on Cycle-MedGAN is described in details. Fig.~\ref{1} depicts an overview of the framework.
\vspace{-2mm}
\subsection{The Cycle-MedGAN framework}
\label{sec:cycleGAN}
The Cycle-MedGAN framework is based on Cycle-GAN \cite{20}, which performs the task of unpaired image-to-image translation from a source domain $X$ (motion-corrupted MR scans) to a target domain $Y$ (motion-free MR scans). The idea behind it is to train a set of two GANs simultaneously. The upper part of the architecture in Fig.~\ref{1} consists of a generator $G_1$ and a discriminator $D_1$, which are trained adversarially to learn the mapping $X\rightarrow{Y}$. $G_1$ attempts to generate images $\hat{y} = G_1(x)$ that resemble images in domain $Y$. On the other hand, $D_1$ acts as a classifier that attempts to label $\hat{y}$ as fake images, while classifying the ground-truth motion-free scans $y$ as real. This adversarial training can be expressed as the following min-max optimization over the adversarial loss function:
\begin{equation}
\setlength\abovedisplayskip{4pt}
\setlength\belowdisplayskip{2pt}
\begin{split}
\min_{G_1} \max_{D_1} \mathcal{L}_{\small\textrm{adv}}(G_1,D_1) = & \; \mathbb{E}_{y} \left[\textrm{log} D_1(y) \right] + \\
& \; \mathbb{E}_{x} \left[\textrm{log} \left( 1 - D_1\left(G_1(x)\right) \right) \right]
\end{split}
\end{equation}
\begin{figure*}[!ht]
	\centering
	
	\includegraphics[width=0.88\textwidth]{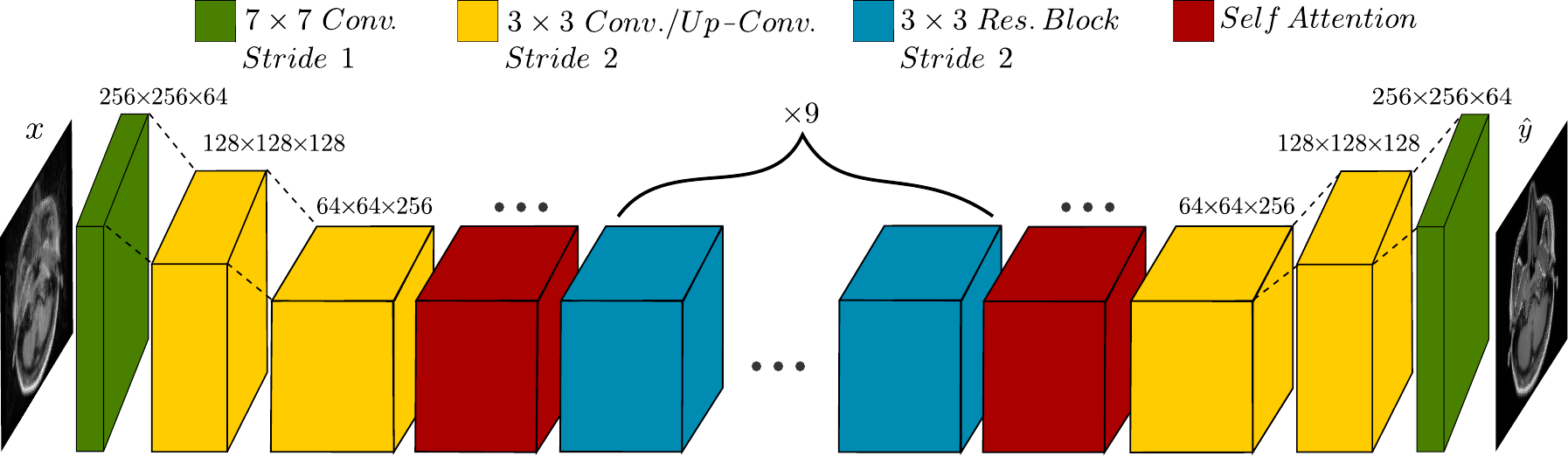}
	
	\caption{The proposed generator architecture with self-attention blocks.}
	\label{2}
	\vspace{-4mm}
\end{figure*}
The lower part in Fig.~\ref{1} consists of generator $G_2$ and discriminator $D_2$, learns the inverse mapping $Y\rightarrow{X}$ in a similar fashion. Moreover, to enable training with unpaired data, the two generators must learn to invert each other, i.e., $\doublehat{x} = G_2(G_1(x)) \approx x$ and  $\doublehat{y} = G_1(G_2(y)) \approx y$.
This is achieved via the pixel-wise cycle-consistency loss:
\begin{equation}
\setlength\abovedisplayskip{5pt}
\mathcal{L}_{\small\textrm{cyc}} = \mathbb{E}_{x} \left[\|x- G_2(G_1(x))\| \right] + \mathbb{E}_{y} \left[\|y-G_1(G_2(y)\| \right]
\end{equation}

However, it has been found out that relying on the pixel loss alone results in blurry and inconsistent results, which fail to capture human visual perception \cite{21}. To alleviate this issue, Cycle-MedGAN introduced two new cyclic feature-based loss functions, namely the \textit{cycle-perceptual loss} and the \textit{cycle-style loss}, to further enhance the perceptual similarity between generated and target images \cite{22}. 
This is achieved by utilizing a pre-trained feature extraction network to learn the feature representation of the input images $x/y$ and their cycle reconstructed counterparts $\doublehat{x}/\doublehat{y}$.
The first loss function, the cycle-perceptual loss ($\mathcal{L}_{\small\textrm{cPercep}}$), aims to capture the high-level perceptual differences between images. With $F_i$ as the activation of the $i^{\textrm{th}}$ layer of a feature extractor network with $L$ layers, this loss can be calculated as the mean absolute error between the feature activations as follows:
\begin{equation}
\setlength\abovedisplayskip{4pt}
\mathcal{L}_{\small\textrm{cPercep}}=\sum_{i = 1}^{L}  \lambda_{cp,i} \left(\|F_i(x)-F_i(\doublehat{x})\|_1+\|F_i(y)-F_i(\doublehat{y})\|_1 \right)
\vspace{-1mm}
\end{equation}
where $\lambda_{cp,i}$ is the weight assigned to the $i^{\textrm{th}}$ layer.
The second loss, the cycle-style loss ($\mathcal{L}_{\small\textrm{cStyle}}$), is inspired by works on neural style transfer \cite{23}. It aims to enhance the style and textural similarity between images. 
This is achieved by first calculating the correlation between different feature maps, which can be 
approximated using the Gram matrices, whose elements are calculated as:
\begin{equation}
\setlength\abovedisplayskip{4pt}
\setlength\belowdisplayskip{4pt}
Gr_i(x)_{m,n} = \frac{1}{h_i w_i d_i} \sum_{h = 1}^{h_i} \sum_{w = 1}^{w_i} F_{i}(x)_{h,w,m} F_{i}(x)_{h,w,n}
\end{equation}
where $h_i{\times}$,$w_i{\times}$ and $d_i$ are the height, width and depth of the 3D feature map $F_{i}(x)$, respectively. This loss is then calculated as the squared Frobenious norm of the difference between the Gram matrices of the inputs and the cycle reconstructed images:
 \begin{equation}
 \setlength\abovedisplayskip{2pt}
 \setlength\belowdisplayskip{2pt}
\begin{split}
\mathcal{L}_{\small\textrm{cStyle}}= & \;  \; \; \, \sum_{i = 1}^{L} \lambda_{cs,i}  \frac{1}{4 d_i^2} \Big( \lVert{Gr_i\left(x\right) - Gr_i\left(\doublehat{x}\right)}\rVert_F^{2}\\
& +   \lVert{Gr_i\left(y\right) - Gr_i\left(\doublehat{y}\right)}\rVert_F^{2} \Big)
\end{split}
\end{equation}
where $\lambda_{cs,i}$ is the weight assigned to the $i^{\textrm{th}}$ layer.
\subsection{MS-SSIM loss}
\vspace{-2mm}
The multi-scale structural similarity index (MS-SSIM) has been widely used in the literature as a metric to evaluate the performance of deep learning models \cite{24}. This is because it takes into account the fact that the human visual system is sensitive to changes in the local structure of an image.
Recently, utilizing MS-SSIM as a training penalty instead of the conventional L1 pixel loss have been reported to result in higher quality images with more detailed local structures \cite{18}. For this purpose, we adapt MS-SSIM as a new cycle loss function to penalize the structural discrepancy between the input images $x/y$ and their cycle reconstructed counterparts $\doublehat{x}/\doublehat{y}$. The proposed cycle MS-SSIM loss is given as \cite{19}:
\begin{equation}
 \setlength\abovedisplayskip{8pt}
\setlength\belowdisplayskip{8pt}
\mathcal{L}_{\small\textrm{MS-SSIM}} =[1-\text{MS-SSIM}(x, \doublehat{x})] + [1-\text{MS-SSIM}(y, \doublehat{y})]
\vspace{-1mm}
\end{equation}
with $\text{MS-SSIM}(x, \doublehat{x})$ as the metric score between $x$ and $\doublehat{x}$.

The total loss function to ensure cycle consistency for our proposed framework is a combination of the four previously mentioned cycle losses. It is given by
\vspace{-1mm}
\begin{equation}
\setlength\abovedisplayskip{8pt}
\mathcal{L}_{\small\textrm{cyc,T}}=\lambda_{L1}\mathcal{L}_{\small\textrm{cyc}}+\lambda_{mS}\mathcal{L}_{\small\textrm{MS-SSIM}}+\lambda_{cP}\mathcal{L}_{\small\textrm{cPercep}}+\lambda_{cS} \mathcal{L}_{\small\textrm{cStyle}}
\end{equation}
where $\lambda_{L1}$, $ \lambda_{\text{\it{mS}}}$, $\lambda_{cP}$ and $\lambda_{cS}$ are the weights assigned for the different loss components respectively. These hyperparameters have been empirically optimized by trial-and-error.
\vspace{-3.5mm}
\subsection{Self Attention Architecture}
\vspace{-2mm}
Similar to Cycle-MedGAN, we utilize a generator network comprising of convolutional layers along with several residual blocks \cite{25}. However, the receptive field of the convolutional layers is local and fails to capture long range dependencies across image regions. Thus, we propose a new generator architecture, inspired from Self-Attention (SA) GANs  \cite{17}. The proposed architecture incorporates two SA blocks which enables long range dependency modelling for the motion correction task. A set of weight matrices, implemented using 1x1 convolutions, are learnt and used to decide the contribution of different regions of the source image in synthesizing any particular region of the target image. 
A detailed representation of the architecture is presented in Fig.~\ref{2}. For discriminators, we utilize the same patch architecture proposed in \cite{20}, with the addition of one SA module.
\begin{table}[!t]
	\caption{Quantitative results\label{t1}}
	\vspace{-2mm}
	\centering
	\setlength\arrayrulewidth{0.08pt}
	\small
	\bgroup
	\def\arraystretch{1.15}
	\resizebox{\columnwidth}{!}{%
		\begin{tabular}{r|cccc}
			
			\hline \hline
			\multirow{2}{*}{Model} & \multicolumn{4}{c}{ MR motion correction}\\ & SSIM & PSNR(dB) & MSE & UQI\\
			\hline
			pix2pix & 0.7904 & 21.61 & 497.75 & 0.5106\\
			Cycle-GAN & 0.8013 & 22.42 & 439.71 & \textbf{0.5494}
			\\
			Cycle-MedGAN & 0.7817 & 23.09 & 409.57 & 0.4119
			\\
			Cycle-MedGAN V2.0 & \textbf{0.8090}& \textbf{23.47} & \textbf{375.01} & 0.5493\\
			\hline
		\end{tabular}
	}
	\egroup
	\vspace{-5mm}
\end{table}
%
\begin{figure*}[!ht]
	\centering
	
	\includegraphics[width=0.86\textwidth]{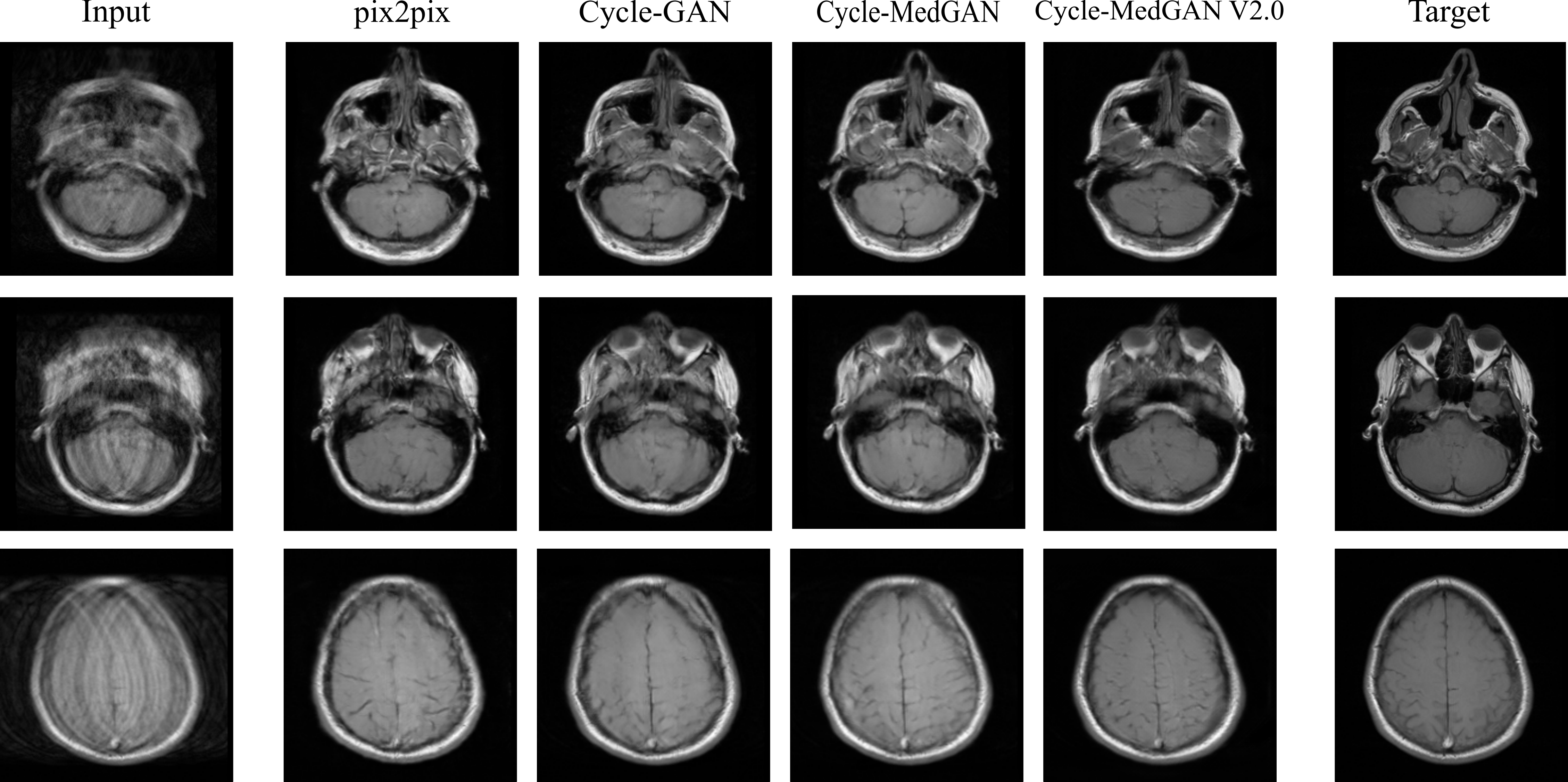}
	
	\caption{Qualitative comparison between our proposed model and other translation techniques for the correction of rigid MR motion artifacts.}
	\label{3}
	\vspace{-2mm}
\end{figure*}
\vspace{-2.5mm}
\section{Experimental Evaluations}
\vspace{0.25mm}
The improved Cycle-MedGAN framework was evaluated for the retrospective correction of severe motion artifacts in brain MRI. For this purpose, a dataset was acquired using a clinical 3 Tesla MR scanner with a T1-weighted spin echo (SE) sequence. For each of the 16 volunteers, two scans were acquired, one under motion-free condition and another where they were asked to freely move their heads. For pre-processing, two-dimensional slices were extracted from the MR volumes and rescaled to $256 \times 256$ pixels. For training, 1101 images from 12 volunteers were utilized while the remaining 436 scans from 4 volunteers were used for validation. The training dataset was explicitly shuffled between the different volunteers to ensure that no input scans (motion-corrupted and motion-free) were paired from the same region or volunteer during the training procedure.

For the feature extractor network, the discriminator of a bidirectional GAN (Bi-GAN) framework \cite{26} pre-trained on a distinct whole-body CT dataset was utilized. We also utilize spectral norm \cite{27} as a discriminator regularizer for improved training stabilization.

To evaluate the performance of our proposed model (Cycle-MedGAN V2.0), we compare it qualitatively and quantitatively against other unpaired image-translation techniques, namely, Cycle-GAN \cite{20} and Cycle-MedGAN \cite{16}.  Also, we compare against the pix2pix framework \cite{28} trained in a supervised manner with no shuffling of the input datasets. This is to investigate the drawbacks of misalignment and non-ideal registration of the input images on the performance of supervised GAN frameworks. All the models used the same network architectures and hyperparameters as recommended in the original publications and were trained for 100 epochs using a single NVIDIA Titan X GPU.

For the quantitative comparisons (Table \ref{t1}), we utilize the Structural Similarity Index (SSIM)
\cite{29}, Peak Signal-to-Noise Ratio (PSNR), Mean Squared Error (MSE) and Universal Quality Index (UQI) \cite{30} as evaluation metrics.
\vspace{-3mm}
\section{Results and Discussion}
\vspace{-1mm}
\label{sec:typestyle}
The quantitative and qualitative results of the proposed framework for unsupervised correction of severe rigid motion artifacts in MR are presented in Fig.~\ref{3} and Table \ref{t1}, respectively. The supervised translation approach, pix2pix, results in worse performance qualitatively and quantitatively when compared with the other unsupervised approaches. Previous works in the literature already reported that misalignment in the input training datasets results in a significant deterioration in the resultant images \cite{15}. In the case of motion artifacts in MR, the imperfections in alignment are due to the separate acquisition of motion-free and motion-corrupted MR data. In contrast, the unsupervised GAN frameworks do not suffer from such a dependency on co-registered datasets. 

 Within the unsupervised approaches, Cycle-GAN results in the worst qualitative performance. Although it succeeds in producing the correct global structure in the resultant brain MR scans, however, Cycle-GAN fails in eliminating the motion blurring effect resulting from severe rigid motion. Our previously proposed Cycle-MedGAN bypasses such problems by utilizing the cycle-style and the cycle-perceptual loss functions to regulate the generator network. This is reflected by the enhanced visual quality of the resultant images. Despite this, traces of motion artifacts can still be observed in the results for severe rigid motion. The approach presented in this work (Cycle-MedGAN V2.0) overcomes this limitation by introducing the SA-based generator architecture and a new MS-SSIM loss function. It exhibits sharper results with fewer traces of motion blurring and better textural details. From a quantitative point-of-view, the same conclusions are also reached by the metrics shown in Table \ref{t1}.

Despite the positive results of the proposed approach, this work is not without limitations. We plan to extend the proposed work for 3-dimensional complex-valued data by additionally incorporating the phase information as a multi-channel input for both rigid and non-rigid artifacts. Additionally, the validity of the diagnostic information can not be guaranteed at this stage. We, therefore, plan to investigate in the future if relevant structures and pathologies are preserved. Thus, at this point, we only consider utilizing such an application for post-processing tasks, e.g. organ volume estimation or semantic segmentation.
\vspace{-3mm}
\section{Conclusion}
\vspace{-0.5mm}
\label{sec:ref}
In this work, we present an improved Cycle-MedGAN framework for the unsupervised retrospective correction of severe rigid motion artifacts. The proposed framework incorporates a new MS-SSIM cycle loss function and a SA-based generator architecture to improve the local structures and to capture long-range dependencies, respectively. Quantitative and qualitative comparisons with other unsupervised translation approaches have illustrated the positive performance of the proposed framework. Additionally, a comparison between the improved Cycle-MedGAN and the supervised translation approach pix2pix have illustrated that the proposed work does not suffer from the same performance deterioration due to training data misalignment. For the future, we plan to conduct more objective evaluations of the results by radiologists.

\newpage
\bibliographystyle{IEEEbib}
\balance
{\footnotesize
	\bibliography{refs}}
\end{document}